# Joint Detections of Frequency and Direction of Arrival in Wideband Based on Programmable Metasurface

He Li, Yun Bo Li, *Member, IEEE*, Wang Sheng Hu, Sheng Jie Huang, Jia Lin Shen, Shi Yu Wang, and Tie Jun Cui, *Fellow, IEEE*

*Abstract* —We propose to achieve joint detections of frequency and direction of arrival in wideband using single sensor based on an active metasurface with programmable transmission states of pass and stop. By integrating two PIN diodes with the opposite directions into the proposed single-layer and ultrathin meta-atom, the transmission performance with 10 dB difference between the pass and stop states is realized in the bandwidth from 5.9 GHz to 8.8 GHz using field-circuit co-simulations. Accordingly, random receiving patterns are generated by controlling the programmable metasurface composed of the switchable meta-atoms. Afterwards, the frequency and direction information of sources located in the far field are detected using the modified algorithm of estimating signal parameters via rotational invariance techniques (ESPRIT) and the compressive sensing method, respectively. A sample of the programmable metasurface is fabricated and the voltage control system is built up correspondingly. To entirely verify the validity of the proposed method, we conduct three kinds of experiments with one single source, double sources with different frequencies, and double sources with the same frequency, respectively. In all cases, the source information of frequency and direction has been detected preciously in measurements in the frequency band from 6.2 GHz to 8.8 GHz.

*Index Terms*—Programmable metasurface, Wideband, Direction of arrival (DoA), Frequency estimation.

## I. Introduction

METASURFACES are composed of subwavelength-scale meta-atoms[1], which have been received much attention in recent years for their powerful abilities to manipulate the electromagnetic (EM) waves and fields. With the advantages of low profile, low cost and easy integration, novel applications have been realized using the metasurfaces, such as hologram generation [2-4], microwave imaging [5, 6], convertion from propagating waves to surface waves [7], superlenses [8, 9] and so on.

To achieve dynamical controls of EM characteristics in real time, promising and challenging researches of programmable metasurfaces [10, 11] were introduced, which can be realized by integrating active elements such as PIN diodes [12], varactor diodes [13] and MEMS [14] into the meta-atoms. Up to now, the reflection-type active metasurfaces with phase or amplitude control [15, 16] have been well explored in the literatures to modulate the scattering waves in the programmable way. Accordingly, to increase the degree of freedom to manipulate the EM waves, space-time coding metasurfaces [17] were proposed by using periodic time modulation of the reflection phase controlled by the bias voltages, which can complete the nonlinear functionalities of harmonic-frequency generation and beam steering simultaneously. More recently, information metasurfaces [18, 19] including the software intelligent types, self-adaptive types and cognitive types have become a new trend by combining the active designs of metasurfaces with the sensing algorithms to realize the joint functions of detections and communications.

Correspondingly, the programmable metasurfaces with transmission phase controls have also been well researched in recent years, and the design methods mainly consist of three types including the electric resonance [20, 21], Hygens' surface [22, 23], and the combined type with receiving antenna (Rx), programmable phase shifter, and transmitting antenna (Tx) [12-14, 24-26]. Compared with the electric resonance structure that suffers from high profile and the Hygens' approach that suffers from ultra-narrow bandwidth, the configurations of the Rx-programmable phase shifter-Tx are more preferred to get dynamic beamforming for the advantages of lower profile and relative larger bandwidth. Based on this method, some high-performance active transmitarrays are achieved of independent transmission amplitude/phase modulations [27], programmable gain control [28], wide angle [29], and wideband design [30, 31]. Considering the effect of feeding blockage under the reflection-type metasurfaces, the transmission types become more attractive in the applications of detection and sensing systems.

Direction of arrival (DoA) estimation is a significant issue for various application areas like wireless communication [32], radar [33] and sonar [34]. Accordingly, the traditional methods

This work was supported in part by National Natural Science Foundation of China (62271146), Defense Industrial Technology Development Program, Foundation Strengthening Program Technology Field Fund (2020-JCJQ-JJ-266), and in part by General Technical Research Project (20201116-0155-001-001). (*Corresponding author: Yun Bo Li and Tie Jun Cui*)

The authors are with the State Key Laboratory of Millimeter Waves, Southeast University, Nanjing 210096, China (e-mail: mozartliyunbo@163.com, tjcui@seu.edu.cn).

of DoA estimation are all concentrated on the phased array including multiple sensors and corresponding signal processing algorithms such as the multiple signal classification (MUSIC) algorithm [35], estimating signal parameters via rotational invariance techniques (ESPRIT) algorithm [36, 37], and some other optimized algorithms [38, 39]. To overcome the disadvantages of high costs by multiple radio-frequency (RF) chains in phased array and the high-complexity signal process algorithms, time-modulated array with switchable RF channel was proposed to detect the directions of the incoming waves, which changes the DoA detection to the frequency-spectrum estimation using harmonic characteristic matrix [40, 41]. The space-time coding metasurface was also presented [42, 43] by using single sensor under the single radio frequency. However, the massive time-domain waveform control should be applied to generate multi-harmonic radiated patterns. Nowadays, the computational imaging systems using compressive sensing method under random receiving patterns that are generated by the programmable or dispersive metasurfaces have become the research highlights [44-47]. Correspondingly, the compressive sensing method can be also applied to the DoA estimation by using the programmable metasuface with only single sensor [48, 49]. In addition, some other signal processing algorithms based on the metasurface architecture like the modified atomic norm method (ANM) were proposed for the new unmanned aerial vehicle (UAV) swarm system [32, 39]. Up to now, the metasurface-based architectures are all concentrated on the direction detections of the incident waves.

The frequency information is also significant and necessary to be detected in the real applications. Therefore, the joint detections of frequency and DoA have received much attention, and various signal processing algorithms based on dimension-reduced methods under multiple RF chains of phased array system are introduced in the previous literatures. In order to estimate the frequency information and DoA simultaneously with low cost and easy integration, the transmission-type programmable metasurface with single sensor can be selected. However, the programmable phase control is hard to realize wideband design under low profile. In this paper, we propose a metasurface with programmable transmission states of pass and stop to accomplish the joint detections of frequency and DoA using the single sensor in wideband. The frequency and DoA information can be detected under the modified ESPRIT algorithm and compressive sensing method, respectively, based on multiple random receiving patterns generated by the programmable metasurface. In our measurement, the joint detections of frequency and DoA information are accurately realized in three representative categories of experiments in wide frequency band from 6.2 to 8.8 GHz (with the relative bandwidth of 34.67%).

The paper is organized as follows. Section II introduces the structure of the proposed meta-atom. The theoretical analyses and simulations of the joint detection algorithms are presented in Section III. Finally, the measurement results of the fabricated programmable metasurface and the estimation results of the three representative categories with one single source, double sources with different frequencies, and double sources with the same frequencies are shown in Section IV.

## II. DESIGN OF UNIT CELL

The joint detections of the frequency and DoA based on the programmable transmission metasurface using single receiver are achieved in this paper, and the schematic diagram is shown in Fig. 1. Firstly, in order to detect the frequency information of the incident wave, the time-domain receiving signals under arbitrary coding apertures are processed using the modified ESPRIT algorithm with multiple time delays [36]. Afterwards, based on the estimated frequency, the corresponding sensing matrix constructed by the far-field patterns under 60 random coding apertures is constructed to acquire the DoA information of the incoming waves.

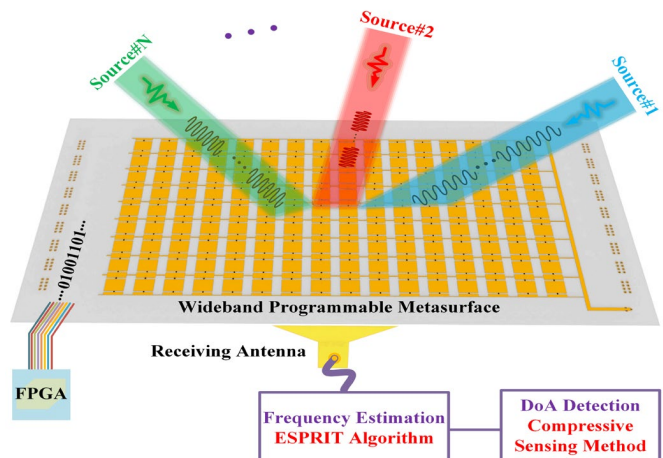

Fig. 1 The schematic diagram of the proposed programmable transmission metasurface to realize the joint detection of the frequency and DoA under the wide band.

To complete the joint detection of the frequency and DoA with single sensor, the wideband programmable metasurface with switchable transmission states of pass and stop is applied, and the texture of the meta-atom is shown in Fig. 2(a). The design mainly consists of two metal layers loaded on the substrate F4B220 with the dielectric constant of 2.2 and the dielectric loss tangent of 0.003. The period size of this unit cell is 15 mm (0.375 $\lambda$ at 7.5 GHz), and the total profile is 1.5 mm (0.0375 $\lambda$ at 7.5 GHz). As shown in Fig. 2(a), the upper layer is composed by square metal patches in *x-axis* direction with two opposite PIN diodes (SMP1320-079LF) integrated, and the EM response of x-polarization is only considered in our design. And the middle patch is connected to the DC bias line in *y-axis* direction fabricated on the bottom layer through the metal vias present in Fig. 2(b). Correspondingly, the other two patches also work as DC ground for PIN diodes. The meta-atom embedded with PIN diodes is co-simulated combining with the commercial softwares of HFSS and ADS, and the schematic diagram of the field-circuit co-simulation is present in Fig. 2(c). And the optimal dimensions of the branch lines for achieving good performance of amplitude difference between pass and stop state are listed in Tab I.

The red dotted line and blue solid line are corresponding to the ON-state and OFF-state of PIN diodes. It can be seen that the transmission amplitude can be changed more than 10 dB difference between the two states from 5.9 GHz to 8.8 GHz with the relative bandwidth of 39.46%.

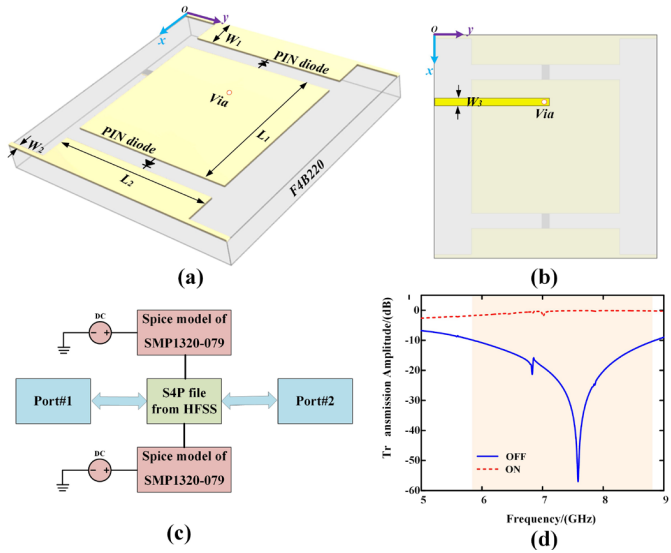

(a) (b) (c) (d)

Fig. 2 (a). The schematic diagram of the proposed meta-atom with the programmable transmission states of pass and stop. (b). The bottom layer of the meta-atom. (c). The schematic diagram of the field-circuit co-simulation method. (d). The simulation results of the transmission amplitude with different states of the PIN diodes under x-polarization.

According to the simulation results shown in Fig. 2(d), the transmission magnitude can be changed by switching the control voltages of the PIN diodes in the wide operating bandwidth. Therefore, the massive radiated patterns with large difference in the far field can have the opportunity to be constructed by whole programmable metasurface with randomly switching the transmission states of the each unit cell under the analogous bandwidth shown in Fig. 2(d).

TABLE I
DIMENSIONS OF BRANCH LINES IN THE META-ATOM

| Parameter | Value(mm) | Parameter | Value(mm) |
|---|---|---|---|
| $W_1$ | 2 | $L_1$ | 9 |
| $W_2$ | 0.25 | $L_2$ | 10 |
| $W_3$ | 0.5 | | |

## III. JOINT DETECTION METHOD OF FREQUENCY AND DOA

The joint detection of frequency and DoA has been separated into two steps. For the first step, we modify the conventional ESPRIT algorithm to estimate the frequency information according to the architecture of programmable metasurface rather than the phased array system. And for the second step, the method of compressive sensing is applied to acquire the DoA information. This joint detection method composed of two steps can be considered as the dimensional reduction which can save time compared with the method of direct two-dimensional data processing.

### A. Frequency Estimation

Considering a programmable metasurface with $M \times N$ unit cells, the receiving signal $x(t)$ at single receiving sensor can be calculated as:

$$x(t)=\sum_{m=1}^{N}\sum_{n=1}^{N}L_{mn}\sum_{i=1}^{K}e^{-j2\pi d\left[(m-1)f_i\sin(\theta_i)\cos(\varphi_i)+(n-1)f_i\sin(\theta_i)\sin(\varphi_i)\right]/c}s_i(t) \quad (1)$$

in which $L_{mn}$ refers to the coding state of the $(m,n)$-th unit cell, $s_i(t)$ is the $i$-th incident wave with the carrier frequency $f_i$, and $(\theta_i,\varphi_i)$ is the direction of arrival, respectively. Compared with the traditional architecture using phased array, only one RF chain is required to receive the output signal. By introducing a time delay $\tau$, the receiving part can be described as:

$$\begin{aligned}x(t-\tau) &= \sum_{m=1}^{M}\sum_{n=1}^{N}L_{mn}\sum_{i}^{K}e^{-j2\pi\left[(m-1)f_i\sin(\theta_i)\cos(\varphi_i)+(n-1)f_i\sin(\theta_i)\sin(\varphi_i)\right]/c}s_i(t-\tau)\\ &=\sum_{m=1}^{M}\sum_{n=1}^{N}L_{mn}\sum_{i}^{K}e^{-j2\pi\left[(m-1)f_i\sin(\theta_i)\cos(\varphi_i)+(n-1)f_i\sin(\theta_i)\sin(\varphi_i)\right]/c}s_i(t)e^{j2\pi f_i\tau}\end{aligned} \quad (2)$$

Correspondingly, the receiving signals with multiple time delays can be denoted as [36]:

$$x(t-n\tau)=LA\Phi^{n-1}S \quad (3)$$

in which the source matrix $S$, direction matrix $A$, and the matrix $\Phi$ generated by the time delay are shown as follows:

$$S=\begin{bmatrix}s_1(t) & s_2(t) & \cdots & s_K(t)\end{bmatrix} \quad (4)$$

$$A=A_x \odot A_y \quad (5)$$

$$\Phi=diag\left\{e^{j\beta_1},e^{j\beta_2},\cdots,e^{j\beta_K}\right\} \quad (6)$$

where $\odot$ refers to the Khatri-Rao product, $\beta_i=2\pi f_i\tau$, and $A_x$ and $A_y$ are denoted as:

$$A_{x(y)}=\begin{bmatrix}1 & 1 & \cdots & 1\\ e^{-j\alpha_{x(y)1}} & e^{-j\alpha_{x(y)2}} & \cdots & e^{-j\alpha_{x(y)K}}\\ \vdots & \vdots & \ddots & \vdots\\ e^{-j(M-1)\alpha_{x(y)1}} & e^{-j(M-1)\alpha_{x(y)2}} & \cdots & e^{-j(M-1)\alpha_{x(y)K}}\end{bmatrix} \quad (7)$$

in which $\alpha_{xi}=2\pi df_i\sin(\theta_i)\cos(\varphi_i)/c$ and $\alpha_{yi}=2\pi df_i\sin(\theta_i)\sin(\varphi_i)/c$. By introducing $P$ time delays, the receiving signals are calculated as:

$$X=\begin{bmatrix}x(t)\\ x(t-\tau)\\ \vdots\\ x(t-P\tau)\end{bmatrix}=\begin{bmatrix}LAS\\ LA\Phi S\\ \vdots\\ LA\Phi^{P-1}S\end{bmatrix} \quad (8)$$

According to the receiving information with multiple time delays, the frequency can be estimated using ESPRIT algorithm. Firstly, the covariance matrix is calculated as $R_x=XX^H$, and the signal subspace $E_s$ is obtained by eigenvalue decomposition, which can be described as:

$$E_s=\begin{bmatrix}LA\\ LA\Phi\\ \vdots\\ LA\Phi^{P-1}\end{bmatrix}T \quad (9)$$

Correspondingly, we can define two matrices $E_1$ and $E_2$, which are given as:

$$E_1 = \begin{bmatrix} LA \\ LA\Phi \\ \vdots \\ LA\Phi^{P-2} \end{bmatrix} T \text{ and } E_2 = \begin{bmatrix} LA\Phi \\ LA\Phi^2 \\ \vdots \\ LA\Phi^{P-1} \end{bmatrix} T \quad (10)$$

And the relationship between $E_1$ and $E_2$ can be seen that:

$$E_2 = \begin{bmatrix} LA \\ LA\Phi \\ \vdots \\ LA\Phi^{P-2} \end{bmatrix} \Phi T = \begin{bmatrix} LA \\ LA\Phi \\ \vdots \\ LA\Phi^{P-2} \end{bmatrix} TT^{-1}\Phi T = E_1 T^{-1}\Phi T \quad (11)$$

According to Eq. (11), we can define $\Psi = E_1^+ E_2$, which has the same eigenvalues with the matrix $\Phi$. Therefore, the frequency information of the *i-th* incident wave can be estimated by the eigenvalue decomposition of the matrix $\Psi$, and the results can be calculated with the eigenvalues $\lambda_i$:

$$\hat{f}_i = \frac{1}{2\pi\tau} angle(\lambda_i) \quad (12)$$

Considering the noise in the free space, the number of the eigenvalues is the same as the dimension of $E_2$. The frequency information is estimated using the $K$ maximum eigenvalues corresponding to the number of the incoming waves. It is noted that these eigenvalues are usually much larger than those generated by noise. However, for the coherent sources, only the maximum eigenvalue refers to the signal subspace. Therefore, we can estimate the frequency information by judging the eigenvalues rather than selecting the $K$ maximum ones directly in our simulations and experiments.

### B. DoA Detection

After estimating the frequency values, the DoA information corresponding to each frequency point can be detected using the compressive sensing method. Similar to the computational imaging system based on the programmable or dispersive metasurface [48, 49], the DoA problem can be solved using the varied coding apertures working as the receiver. Firstly, the sensing matrix is constructed with the far-field patterns generated by the random transmission amplitude distributions of the programmable metasurface. The corresponding radiation pattern $F(\theta,\varphi)$ can be calculated by using the 2D Fourier transform from the near field to the far field:

$$\vec{F}(\theta,\varphi) = j2\pi k \cos(\theta)\vec{A}(k\sin(\theta)\cos(\varphi),k\sin(\theta)\sin(\varphi)) \quad (13)$$

$$\begin{cases} A_x(\vec{k}) = \frac{e^{jk_z d}}{4\pi} \int_{-\infty}^{+\infty} E_x(x,y,d) e^{j(k_x x + k_y y)} dxdy \\ A_y(\vec{k}) = \frac{e^{jk_z d}}{4\pi} \int_{-\infty}^{+\infty} E_y(x,y,d) e^{j(k_x x + k_y y)} dxdy \end{cases} \quad (14)$$

in which $A_x$ and $A_y$ refer to the Fourier transform results from the transverse electronic fields $E_x$ and $E_y$. Accordingly, the sensing matrix $G$ under each frequency point can be described as:

$$G = [F_1(\theta,\varphi,f_i), F_2(\theta,\varphi,f_i), \cdots, F_L(\theta,\varphi,f_i)]^T \quad (15)$$

in which $L$ means the number of the random coding apertures, and $f_i$ refers to the frequency of the *i-th* incident waves.

Afterwards, when the sources are excited in the far field, the receiving amplitude and phase vector $E_R \in C^{L \times 1}$ corresponding to the varied coding apertures are acquired, which is described as:

$$E_R = GS + N \quad (16)$$

where $S \in C^{L \times M}$ is a row vector representing the directions of the incident waves, and the far-field area is divided into $M$ parts. $N$ refers to the noise matrix in the free space. It is noted that the directions of incident waves can be perfectly solved when $G$ is a non-singular matrix. However, considering underdetermined situation of $L<M$, the compressive sensing method should be used to solve the ill-posed problem for the estimation of $S$.

Accordingly, the Tikhonov regularization method is applied to replace the ill-posed problem with well-posed problem by introducing the penalty term $\lambda$. Therefore, the estimated result can be obtained as:

$$\hat{S} = (G^H G + \lambda G)^{-1} G^H E_R \quad (17)$$

The selection of penalty term $\lambda$ is hand-tuned for one of the DoA detection and this value is used for all situations.

### C. Simulation Results

According to the modified ESPRIT algorithm and the compressive sensing method proposed in the previous parts, the joint detections of the frequency and DoA information is verified in commercial software MATLAB. The programmable metasurface is composed of $16 \times 9$ unit cells in our calculations. Firstly, the frequency information of the coming wave is estimated by applying the modified ESPRIT method. Then the elements of $G$ matrix under the estimated frequency can be constructed with the data of random radiation patterns generated by the programmable apertures shown in Fig. 3. Thus the DoA information can be acquired according to Eq. (17). Three representative categories of DoA estimations including one single source, double sources with the same frequency, and double sources with different frequencies are preferred and the corresponding calculation results are shown in Fig. 4.

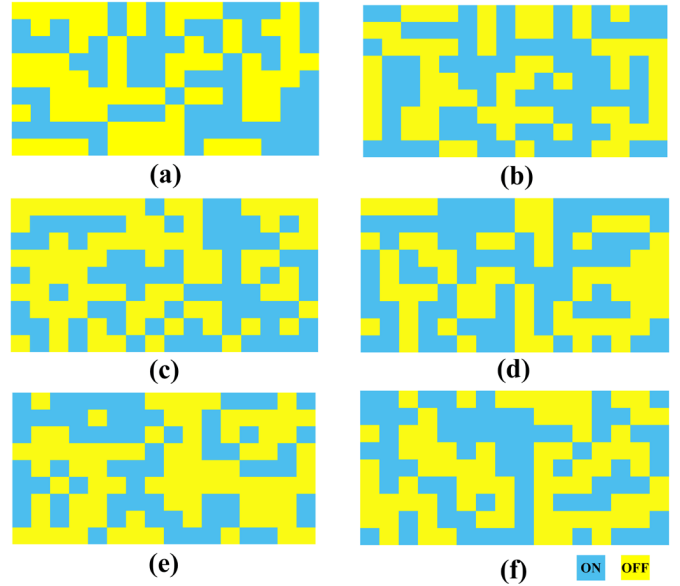

Fig. 3 Several typical random coding distributions of the programmable metasurface composed of the unit cells with switchable transmission amplitude of pass and stop.

In our calculations, the sources located in the azimuth plane is only considered in order to make comparison with the further measurement, which is hard to test in the two-dimensional space. The frequency points at band boundaries (5.9 GHz, 8.8 GHz) and a middle frequency point (7.3 GHz) in the band are selected to verify the proposed design. The results shown in Fig. 4 demonstrate that the frequency estimation can achieve the performance of high accuracy with the error lower than 0.02 GHz, corresponding to a relative error of 0.27%. According to the estimated frequency point, the corresponding $G$ matrix constructed by 60 random far-field radiation patterns (from -90° to 90° with 1° interval) is introduced to detect DoA using the Tikhonov regularization method. The calculated results of three selected categories of one single source (shown in Fig. 4(a)-(c)), double sources with different frequencies (Fig. 4(d)-(f)) and double sources with the same frequency (Fig. 4(g)-(i)) have the good consistencies to the predesigned angles marked by the green dashed lines in Fig. 4.

## IV. Experimental Results

### A. Implementation of the Programmable Metasurface

A prototype of the programmable metasurface consisting of 16×9 meta-atoms is fabricated. The upper layer shown in Fig. 5(a) presents that two PIN diodes are set with opposite directions and a metal via is designed to connect the bottom layer working as the bias line which is shown in Fig. 5(b). Each unit cell can be independently controlled by external voltage module of NI PXIe-6739R, and the DC signals are isolated with RF ones by using the inductance of 1000 nH.

To verify the performance of the switchable transmission amplitudes, the fabricated sample is calibrated under two horns and the schematic diagram of the measurement setup is shown in Fig. 6(a). By setting the voltages of all the PIN diodes with the same states, the transmission amplitude between two horns are acquired with vector network analyzer (VNA) of Keysight N9928A. The calibrated results are present in Fig. 6(b), from which we note that the transmission amplitude can reach 10 dB difference between the two states of ON and OFF under the frequency band from 6.2 GHz to 8.8 GHz with the relative bandwidth of 34.67%. The results shown in Fig. 6(b) have a little deviation from the co-simulations, which may be caused by the finite array, horn excitation, and fabricated error.

### B. Joint Detections of the Frequency and DoA

Accordingly, the fabricated programmable metasurface is set in a microwave anechoic chamber to realize the joint detections of the frequency information and DoA, and the experiment setup is shown in Fig. 7. The incident sources are located in the far-field region of the programmable metasurface, and a single horn antenna is working as the sensor to receive the signals with varied coding patterns generated by the metasurface.

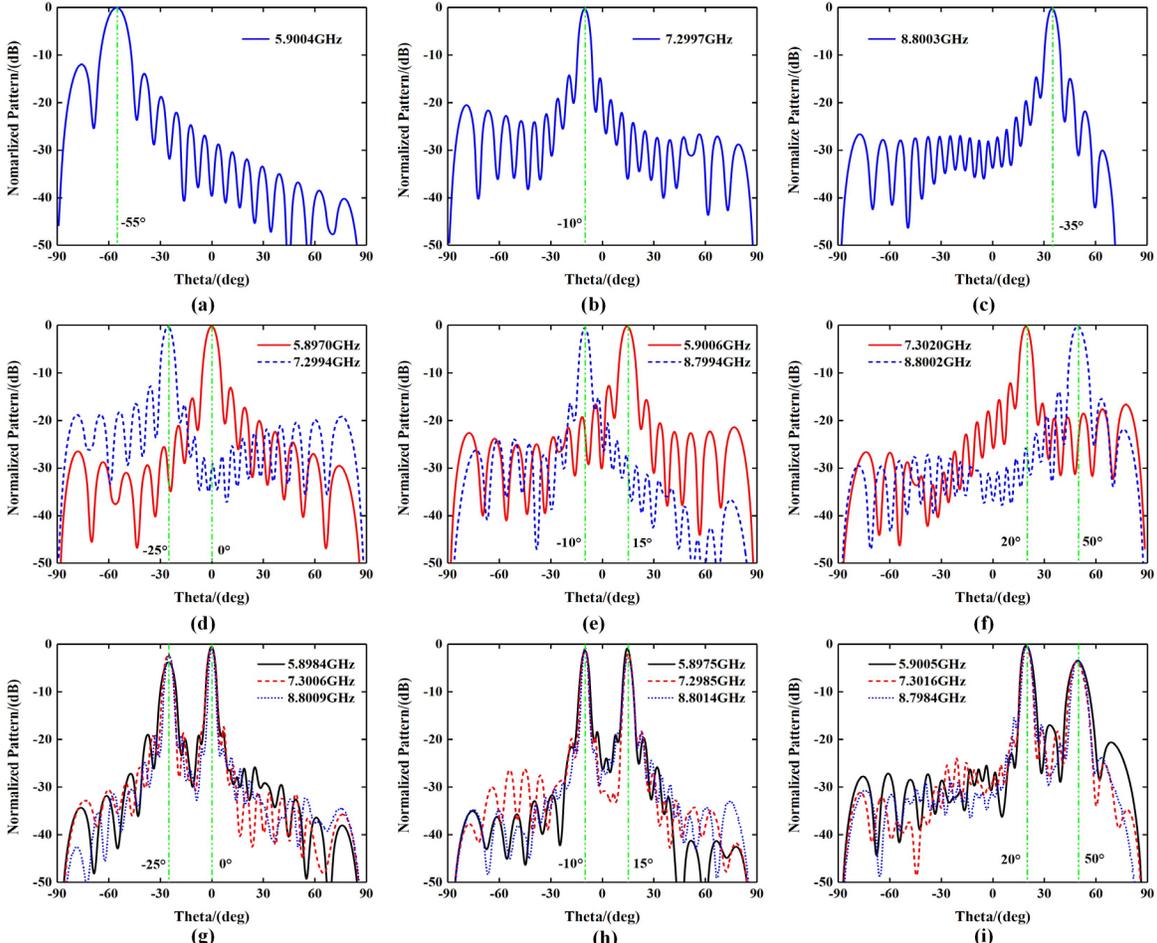

Fig. 4 The frequency estimation and DoA detection results of (a-c). one single source, (d-f). double sources with different frequencies and (g-i). double sources with the same frequency in our calculations.

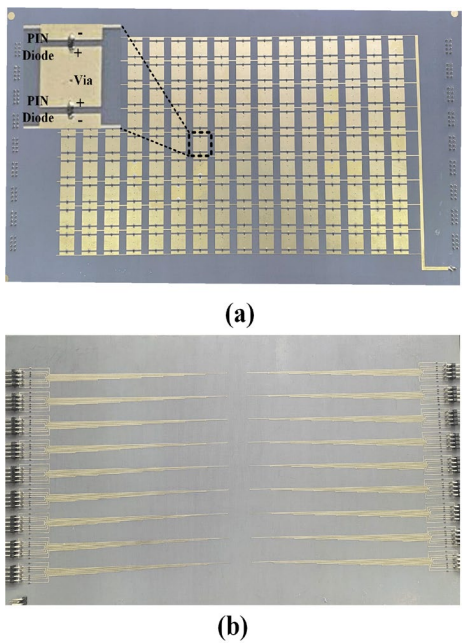

Fig. 5 Photographs of the fabricated programmable transmission metasurface. (a). The upper layer. (b). The bottom layer.

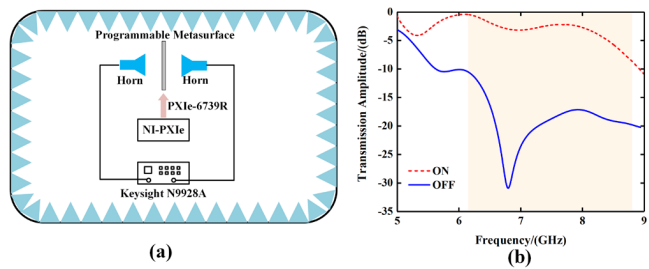

Fig. 6 (a). The schematic diagram of the measurement setup for calibration. (b). Measured results of switchable transmission amplitude.

Firstly, by rotating the fabricated sample located on the platform from –90° to 90° with 1° interval, the amplitudes and phases of the far-field patterns corresponding to the 60 random coding patterns are acquired under the frequency points from 6 GHz to 9 GHz with 0.015GHz interval using VNA (Keysight N9928A). The calibrated results are obtained to construct the sensing matrix, and some of the measurement far-field patterns under the frequency points of 6.2 GHz, 7.5 GHz and 8.8 GHz are presented in Fig. 8. The measured results indicate that the patterns are different enough so that they can be applied for the DoA detection using the compressive sensing method.

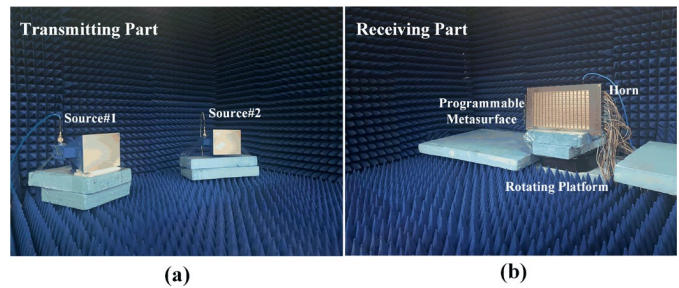

Fig. 7 The experiment setup in the microwave anechoic chamber. (a) The setup of the sources; (b) the setup of the fabricated programmable metasurface.

Afterwards, three representative categories of experiments, one single source, double sources with different frequencies, and double sources with the same frequency, are placed in the far-field region. The sources are connected to a signal generator (Agilent E8257D). With the modified ESPRIT algorithm, the frequency information is firstly estimated by receiving the time domain signals using the oscilloscope (Agilent DS091304A). Then, the corresponding sensing matrix is obtained to detect the DoA with the Tikhonov regularization method by receiving the frequency-domain signals using the spectrum analyzer (Agilent E4447A).

The measurement results of joint detections of frequency and DoA are presented in Fig. 9. The edge frequency points of 6.2 GHz and 8.8 GHz, and the central frequency point of 7.5 GHz are selected in the experiments. For the results of one single source shown in Fig. 9(a)-(c) and double sources with different frequencies demonstrated in Fig. 9(d)-(f), only the amplitude information of the far-field patterns are used to reconstruct the directions of the incident waves.

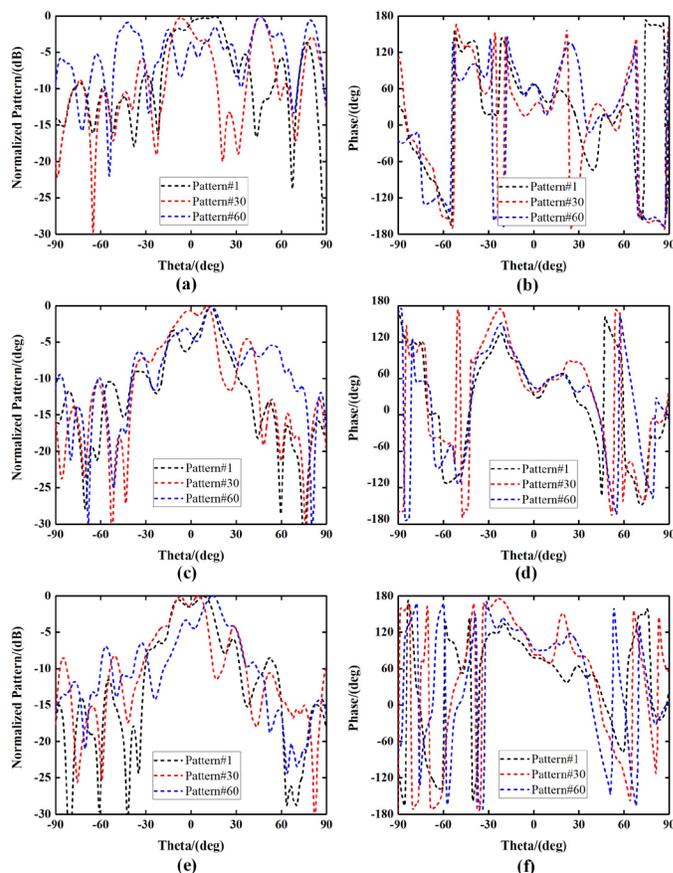

Fig. 8 The measured amplitude and phase of the far-field patterns under the frequency point (a-b). 6.2 GHz, (c-d).7.5 GHz and (e-f). 8.8 GHz.

The estimated frequencies are listed at the legends in the figures, which present very good performance with the relative error lower than 0.2%. The locations of the maximum values under blue and red lines in the results show great consistency to the preset incoming angles marked by the green lines. In addition, double sources with the same frequency are also detected in the experiments and the measured results are shown in Fig. 9(g)-(i). We note that all measured results have very good consistency with the preset source information, which can verify the validity of wideband detections of frequency and

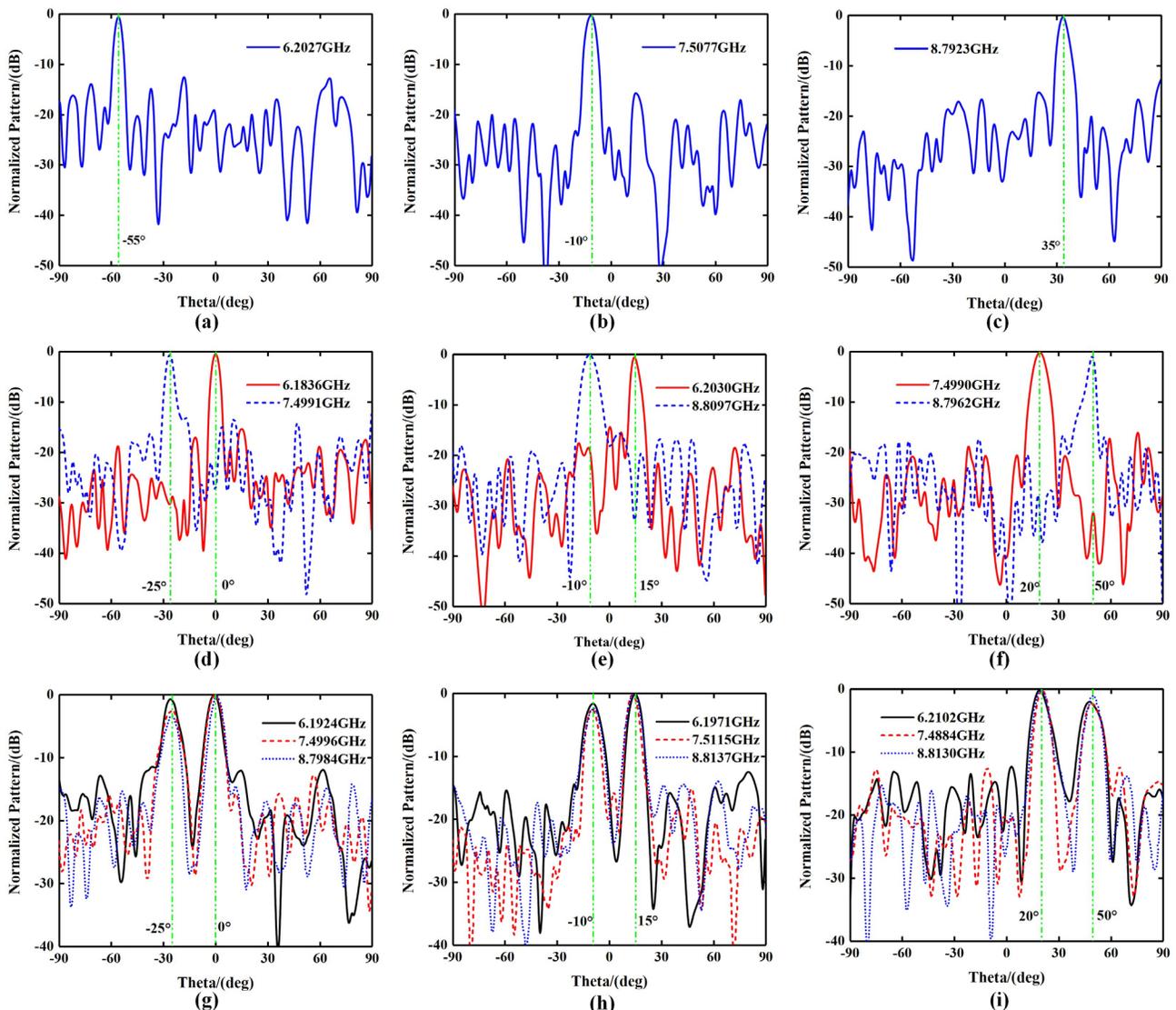

Fig. 9 The measured results of frequency estimations and DoA detections under different scenarios. (a-c). One single source. (d-f). Double sources with different frequencies. (g-i). Double sources with the same frequency in the experiment.

DoA information of the coming waves based on the proposed programmable metasurface

## V. Conclusion

We proposed a new method to realize the wideband detection of frequency and DoA information of the coming waves based on the programmable metasurface composed of meta-atoms with switchable transmission states of pass and stop. The ultra-thin (0.0375 λ at 7.5GHz) unit cell is designed by integrating two opposite PIN diodes with only single layer. Afterwards, the programmable transmission metasurface of 16×9 unit cells is fabricated and measured, and the calibrated results show that the transmission performance of 10 dB difference can be achieved within the frequency band from 6.2 GHz to 8.8 GHz, which shows good consistency to the co-simulation results of the bandwidth from 5.9 GHz to 8.8 GHz.

Accordingly, the joint detection of frequency and DoA is realized using proposed programmable metasurface with single sensor in the experiment. The frequency and DoA of the sources are estimated by modified ESPRIT method with time-domain signals and compressive sensing method with frequency-domain signals respectively. For measurement results, the good consistency with preset source information is observed that can verify the validity of wideband detection of frequency and DoA information based on proposed programmable metasurface. Due to the advanced features of wideband and ultra-low-profile design, the presented new sensing technology can have the huge potential to be introduced in the modern radar and communication systems.